\documentclass[fleqn]{article}

\newcommand{\titel}
{A progression ring for interfaces of instruction sequences,
threads, and services
\thanks{The authors acknowledge support from
 the NWO project Thread Algebra for Strategic Interleaving.}
}

\usepackage{stmaryrd,amssymb,amsmath,amsthm,url}

\newtheorem{definition}{Definition}  
\newtheorem{example}{Example}
\theoremstyle{definition}

\newcommand{\betaw}{\beta}
\newcommand{\length}{\ell}
\newcommand{\dd}[1]{\displaystyle{\frac\partial{\partial #1}}}
\newcommand{\ddm}[1]{\displaystyle{\frac{\overline\partial}{\partial #1}}}

\newcommand{\jum}[1]{\ensuremath{\##1}}
\newcommand{\bjum}[1]{\ensuremath{\backslash\##1}}
\newcommand{\ter}{\ensuremath{\:!\:}}

\newcommand{\di}{\mathsf{D}}
\newcommand{\st}{\mathsf{S}}

\newcommand{\NI}{{\mathbb I}}

\newcommand{\Nat}{{\mathbb N}}

\newcommand{\tr}{\ensuremath{\mathtt{true}}}
\newcommand{\fa}{\ensuremath{\mathtt{false}}}

\title{\titel}

\author{
	Jan A.\ Bergstra \and
	Alban Ponse \\
\\
  {\small
	  Section Software Engineering,
	  Informatics Institute,
	  University of Amsterdam}\\
	{\small URL: \url{www.science.uva.nl/~{janb/, alban/}}
	}
}
\date{}

\begin{document}
\maketitle

\begin{abstract}
We define focus-method interfaces and some 
connections between such
interfaces and instruction sequences, giving rise to
instruction sequence components.
We provide a flexible and practical notation for interfaces
using an abstract datatype specification comparable to that
of basic process algebra with deadlock. The structures
thus defined are called progression rings.
We also define thread and service components.
Two types of composition of instruction sequences 
or threads and 
services (called `use' and `apply') are lifted to 
the level of components. 
\end{abstract}

\section{Introduction}
\label{sec:intro}
We can not simply say that this instruction sequence
(\emph{inseq})
has that interface because there are different ways
to obtain an interface from an inseq, which is one the
main questions that we deal with in this paper.
Instead we will do this:
\begin{itemize}
\item[$(i)$]
we define and formalize so-called focus-method 
interfaces, or briefly, FMIs,
\item[$(ii)$]
we specify some relations between FMIs and inseq's, and
\item[$(iii)$]
we define an \emph{inseq component} as a pair $(i,P)$ of an
FMI $i$ and an inseq $P$ where $i$ and $P$ need to be
related in the sense meant in $(ii)$ above.
\end{itemize}

\noindent
Focus method interfaces will also serve as thread-component 
interfaces, whereas MIs (method interfaces) will be used
as service-component interfaces.

Like with an instruction sequence, a \emph{thread component}
results by pairing a thread and an FMI, 
and a service component consists of a pair of a 
service and an MI.

\bigskip

Using \emph{focus-method notation} (FMN) a 
\emph{basic instruction} has the form
\begin{align*}\texttt{f.m}
\end{align*}
where $\texttt f$ is a \emph{focus} and $\texttt m$ 
is a \emph{method}. 
Furthermore, we have
the following test instructions:
\begin{align*}
+\texttt{f.m}&\text{ (a \emph{positive test instruction})}\\
-\texttt{f.m}&\text{ (a \emph{negative test instruction})}
\end{align*}
Some examples of instructions in FMN are:
\[\texttt{b17.set:true}\quad\text{and}\quad
{+}\texttt{b18a.get}\]
where \texttt{b17} and \texttt{b18a} are foci addressing 
certain boolean registers, and \texttt{set:true} and
\texttt{get} are methods defined for such boolean registers.

The idea of a positive test instruction is that upon
execution it yields a reply \tr\ or \fa, and that upon
the reply \tr\ execution continues with the next instruction,
while upon reply \fa\ the next instruction is skipped and
execution continues with the instruction thereafter.
The execution of a negative test instruction 
displays
the reversed behavior with respect to the reply values.
Note that the reply values \tr\ and \fa\ have nothing to do
with the example method \texttt{set:true} mentioned above.

In \cite{BL02} the programming notation PGLB is defined:
Next to a given set 
$A$ of basic instructions
and the test instructions generated from $A$, PGLB contains
forward jumps $\jum k$ and backward
jumps $\bjum k$ for $k\in\Nat$ as primitive instructions,
and a termination instruction which is written as $\ter$. The 
instructions mentioned here are PGLB's so-called 
\emph{primitive} instructions. The inseqs that we call
PGLB programs are formed from primitive instructions
using concatenation, notation $;$. 
The following three inseqs are examples of PGLB programs:
\[\begin{array}{l}
\texttt{b17.set:true}\\[2mm]
+\texttt{b18a.get};\jum 2;
\texttt{b17.set:true}\\[2mm]
-\texttt{b1.get};\jum3;\texttt{b4.set:false};\jum3;
\texttt{b4.set:true};\ter\\
\end{array}\]
We consider PGLB as an inseq notation, and we
assume that adaptations to other inseq notations 
can easily be made.

In \cite{BL02} it is defined in what way a PGLB program
defines a thread. Here we now discuss the interface of 
such inseqs.
Consider 
\[P=-\texttt{b1.get};\jum3;\texttt{b4.set:false};\jum3;
\texttt{b4.set:true};\ter\]
It is reasonable to view
\[
\{\texttt{b1.get},\texttt{b4.set:false},\texttt{b4.set:true}\}
\]
as the interface of $P$. We will call this a focus-method
interface (FMI) for $P$ as it consists of focus-method pairs.

For an interface it is important that its description 
is ``simple''. If an interface of inseq $P$ is automatically 
derived from $P$ then storage of that interface as a part 
of the code is a matter of pre-computation.

These pre-computed datastructures should be easy to 
store and easy to use. This ``use'' can be a matter of
application of static checks which prevent the occurrence
of dynamic errors. It follows that one needs a notation
for interfaces which allows notational simplification.

Writing + for union and omitting brackets for single 
elements
one may write
\begin{align}
\nonumber
\{\texttt{b1.get}, \texttt{b4.set:false},\texttt{b4.set:true}\}
&=\texttt{b1.get}+\texttt{b4.set:false}+\texttt{b4.set:true}\\
\nonumber
&=\texttt{b1.get}+\texttt{b4.set:}(\texttt{false}+\texttt{true})\\
\label{a}
&=\texttt b(\texttt{1.get}+\texttt{4.set:}(\texttt{fals}+\texttt{tru})\texttt e)
\end{align}
We find that this notation allows useful simplifications. 
The main contents of this paper is to specify the details
of this particular interface notation.

We notice that if $i$ is a plausible interface for inseq $P$,
it may be the case that some $i'$ extending $i$ can be denoted
with a significantly more simple FMI-expression, say $i_e$.
Then $i_e$ is also a plausible interface for $P$. We will
provide technical definitions of various forms of matching
inseq's with interfaces in such a way that this intuition 
can be made formal.

From a mathematical point of view the task to provide the
details of the interface notation as suggested in \eqref{a}
is not at all challenging. From the point of view of abstract
data type specification it requires the explicit resolution
of several design alternatives, 
each of which have advantages and 
disadvantages. In particular, insisting that interface elements
are in FMN, one needs to resolve some language/meta-language
issues in order to provide an unambiguous semantics of
interface expressions.

We hold that from some stage onwards the theory of instruction
sequences needs to make use of a pragmatic theory of interfaces.
It is quite hard to point out exactly when and where having
full details of an interface notation available matters.
For that reason we have chosen to view the design of an
interface notation for instruction sequences as an
independent problem preferably not to be discussed 
with a specific application in mind.

Summing up, this is the question posed and answered below:
\begin{quote}
Assuming that instruction sequence interfaces are important
per se, provide a flexible and practical notation for
interfaces by means of an appropriate abstract datatype
specification.
\end{quote}

\bigskip

The paper is structured as follows: In the next section we 
introduce progression rings and show how they underly
an abstract datatype specification for interfaces.
In Section~\ref{sec:req} we define inseq components and
thread components, and in Section~\ref{sec:serv} we discuss
composition of such components with service
components.
In Section~\ref{sec:conc} we 
briefly discuss interfaces that are not minimal 
and make a remark about related work.

\section{Progression Rings and Interfaces}
We first formally define the \emph{letters,  lower case} and 
\emph{upper case} in BNF-notation 
(enclosing terminals in double quotes):
\begin{align*}
V_{LLC}&::=\text{``\texttt a"}\mid \text{``\texttt b''}
\mid \text{``\texttt c''}\mid ...\mid \text{``\texttt z''}~~~
(\text{26 elements})\\
V_{LUC}&::=\text{``\texttt A''}\mid \text{``\texttt \texttt B''}
\mid \text{``\texttt C''}\mid ...\mid \text{``\texttt Z''}~~~
(\text{26 elements})\\
V_L&::=V_{LLC}\mid V_{LUC}
\end{align*}
Furthermore, we shall use digits ($V_D$), the colon and 
the period as terminals for identifiers:
\begin{align*}
V_D&::=\text{``\texttt 0''}\mid ...\mid\text{``\texttt 9''}\\
V_C&::= \text{``\texttt :''}\\
V_{LDC}&::=V_L\mid V_D\mid V_C\\
V_P&::={``\texttt ."}\\
V_{LDCP}&::= V_{LDC}\mid V_P
\end{align*}

Let $I$ be the set of interfaces. Elements of $V_{LDC}$
and $V_P$ (thus of $V_{LDCP}$) are considered constants 
for $I$, and
\[\delta\]
is a special constant denoting the empty interface. 
Furthermore,
 $X,Y,Z,...$ are variables ranging over $I$.

On $I$ we define $+$ and $\cdot$ as alternative and 
sequential composition,
where $\cdot$ binds stronger than + and will be omitted
whenever possible. 
As usual, we use the notation $V^+$ for the set of
finite strings over alphabet $V$.

Consider a string, say
\(\betaw= \texttt{a7:b.b25:8.c}\)
in $(V_{LDCP})^+$. The string $\betaw$ is understood 
as a selector path
to be read from left to right. Stated in different terms,
$\betaw$ contains a progression of information separated
by periods (``\texttt .") to be understood progressively from
left to right.
For this reason we will call $\betaw$ a progression. The
notion of a progression thus emerging combines that of a 
string or word and that of a process.
A process in a bisimulation model can be considered a 
branching progression where progression is thought 
in terms of time or less abstract in terms of causality.

Having recognized interfaces as progressions it is a 
straightforward decision to take an existing algebraic
structure of progressions as the basis of their formal 
specification. To this end we provide the axioms
of Basic Process Algebra with $\delta$, or briefly,
BPA$_\delta$ (see \cite{BW90,F00}) in Table~\ref{tab:bpad}.

\begin{table}[htbp]
\caption{The axioms of BPA$_\delta$}
\vspace{1mm}\hrule
\begin{align*}
X+Y&=Y+X&(X+Y)\cdot Z&=X\cdot Y+X\cdot Z&X+\delta&=X&\\
(X+Y)+Z&=X+(Y+Z)&(X\cdot Y)\cdot Z&=X\cdot (Y\cdot Z)&\delta\cdot X&=\delta\\
X+X&=X
\end{align*}
\hrule
\label{tab:bpad}
\end{table}

\begin{definition}
A \textbf{right progression ring}
is a structure that satisfies 
the equations of BPA$_\delta$
(see Table~\ref{tab:bpad}).
\end{definition}

Thus sequential composition $\cdot$ is right distributive 
over $+$ in a right progression ring and $\delta$ is its
additive identity. Furthermore, $\delta$ is a left zero
element for sequential composition.

For the sake of completeness we also define left
progression rings.
\begin{definition}
A \textbf{left progression ring} 
is a non-commutative ring with respect to
sequential composition, in which 
$+$ is commutative, associative and idempotent, 
and sequential composition $\cdot$ is left distributive 
over $+$. Furthermore there is a constant $\delta$ that 
is both the additive identity and a right zero element for 
sequential composition.
\end{definition}

So a left progression ring satisfies the axioms
\begin{align*}
X\cdot (Y+Z)=X\cdot Y+X\cdot Z&\quad\text{instead of}\quad
(X+Y)\cdot Z=X\cdot Y+X\cdot Z,\\
X\cdot \delta=\delta&\quad\text{instead of}\quad\delta\cdot X=\delta.
\end{align*}
A right or left progression ring is
\emph{distributive} if $\cdot$ is distributive 
over $+$. 

\bigskip

$\NI_{LDCP}$ is the initial distributive right
progression ring with constants taken from $V_{LDCP}$.
In this paper we will only consider $\NI_{LDCP}$ as a
structure for interfaces. 
We will omit the symbol $\cdot$ in a sequential composition
whenever reasonable. In the case that we consider closed terms
that are built with sequential composition
only, we will certainly omit
this symbol, and for example write
\[\texttt{b114.get}\quad\text{instead of for example}\quad
\texttt{b1}\cdot\texttt{14}\cdot\texttt{.get}\]
although these two expressions of course
represent the same closed term.
Observe that when adopting this convention, 
each $\betaw\in (V_{LDCP})^+$
can be seen as an element (an interface, a closed term) in 
$\NI_{LDCP}$. 

We notice that most process algebras as found in \cite{BK84,BW90,F00}
are (non-distributive) right progression rings, some unital 
(i.e., containing a unit for sequential composition).

In $\NI_{LDCP}$ we write 
\[X\sqsubseteq Y\quad\text{if}\quad X+Y=Y.\]
For each $\betaw\in (V_{LDCP})^+$ we need two additional
operators on $\NI_{LDCP}$:
\begin{align*}
\dd\betaw(X)&\quad\text{the $\betaw$-derivative of $X$:}\\
&\quad\text{yields the largest $Y$ such 
that either $Y=\delta$
or $X+\betaw\cdot Y\sqsubseteq X$,}\\[3mm]
\ddm\betaw(X)&\quad\text{the $\betaw$-filter-complement of $X$:}\\
&\quad\text{removes all progressions from 
$X$ which have $\betaw$ 
as an initial segment.}
\end{align*}
For example,
\[\dd{\texttt{a.b}}(\texttt{a.}+\texttt{a.b}+\texttt{a.bc})=
\texttt c=\dd{\texttt{a.b}}(\texttt{a.bc})\]
and
\[
\ddm{\texttt{a.b}}(\texttt{a.}+\texttt{a.b}+\texttt{a.bc})=
\texttt{a.}=\ddm{\texttt{a.b}}(\texttt{a.}).\]

\bigskip

Both these operators can easily be defined.
Let $u,v\in V_{LDCP}$ and $\betaw\in (V_{LDCP})^+$, then
the $\betaw$-derivative $\dd{\betaw}(\_)$
 is defined as follows:
\begin{align*}
\dd u(\delta)&=\delta,
&\dd{u\betaw}(X)=\dd \betaw(\dd u(X)),
\\[2mm]
\dd u(X+Y)&=\dd u(X)+\dd u(Y),
\\[2mm]
\dd u(v)&=\delta,
\\[2mm]
\dd u(vX)&=\begin{cases}
\delta&\quad \text{if $v\neq u$},\\
X&\quad\text{otherwise},
\end{cases}
\end{align*}
and the $\betaw$-filter-complement $\ddm{\betaw}(\_)$
is defined by:
\begin{align*}
\ddm u(\delta)&=\delta,&
\ddm{u\betaw}(\delta)&=\delta,
\\[2mm]
\ddm u(X+Y)&=\ddm u(X)+\ddm u(Y),&\ddm {u\betaw}(X+Y)&=\ddm {u\betaw}(X)+\ddm {u\betaw}(Y),
\\[2mm]
\ddm u(v)&=\begin{cases}
v&\quad \text{if $v\neq u$},
\\[2mm]
\delta&\quad\text{otherwise},
\end{cases}&\ddm {u\betaw}(v)&=
v,\\
\ddm u(vX)&=\begin{cases}
vX&\quad \text{if $v\neq u$},\\
\delta&\quad\text{otherwise},
\end{cases}&\ddm {u\betaw}(vX)&=\begin{cases}
vX&\quad\text{if either $v\neq u$, or $v=u$}\\
&\quad\text{and $\ddm \betaw(X)\neq\delta$},\\
\delta&\quad\text{if $v=u$ and $\ddm \betaw(X)=\delta$}.
\end{cases}
\end{align*}

\section{Instruction Sequence and Thread Components}
\label{sec:req}
A focus-method interface (FMI) is an interface which
is provably equal to either $\delta$ or a term of
the form
\[+_{i=1}^{k}\beta_i\texttt.+_{j=1}^{k_i}\beta_{i,j}\]
with $\beta_i, \beta_{i,j}\in V_L(V_{LDC})^+$ and 
$k, k_{i,j}>0$.
Informally stated this means that after flattening
(bringing + to the outside), we have words with exactly
one period each and no words ending in $\delta$.

Let $i\in\NI_{LDCP}$ be an expression denoting
some FMI (thus $i\neq\delta$).
Then PGLB$_i$ is the instruction sequence notation 
with basic instructions taken from $i$. The number of 
possible basic instructions may grow exponentially
with the size of the expression $i$. 

\begin{example}\rm
\label{int}
Consider the interface $i$ defined by
\begin{eqnarray}
\nonumber
i&=&
\texttt{f.}(\texttt{get}+(\texttt{set:}+\texttt{testeq:})
(\texttt 0+\texttt 1+\texttt 2)(\texttt 0+\texttt 1+\texttt 2)(\texttt 0+\texttt 1+\texttt 2))~+\\
\label{i}
&&\texttt{g.}(\texttt{get}+(\texttt{set}+\texttt{testeq})
(\texttt{:})(\texttt{true}+\texttt{false}+\texttt{error}))
\end{eqnarray}
which abbreviates $1+2\cdot3^3+7=62$ basic instructions (and the 
availability of twice as much test instructions in PGLB$_i$).
\end{example}

Here we notice a first reward of our formalization:
a flexible specification format for a wide variety
of different instruction sequence notations. Instead 
of PGLB one might of course use other other 
program notations such as PGA, PGLC or PGLD from
\cite{BL02}, PGLA from \cite{BP08}, or $C$ from \cite{BP09}.

In \cite{BP02} a `program component' $[i,P]$
is defined as a pair of an interface and a program
with the requirement that $i$ contains at least all
instruction names that occur in program $P$.
We stick to this idea and 
hold that an \emph{instruction sequence component} is a 
pair 
\[(i,P)\]
of a focus-method interface
$i$ and an instruction sequence $P$, where some form
of match between $i$ and $P$ needs to be assumed.
We have a number of options which we will formally 
distinguish:
\begin{definition}
Let $P$ be some inseq in FMN.
Then
 $P$ \textbf{requires}
interface $i$ if each element 
$\textup{\texttt{f.m}}$ of $i$
occurs in $P$ and conversely, all basic instructions 
occurring in $P$ are elements of $i$, where 
$\textup{\texttt{f.m}}$
occurs in $P$ if at least one of $\textup{\texttt{f.m}}, 
+\textup{\texttt{f.m}}, 
-\textup{\texttt{f.m}}$
is an instruction of $P$.

Furthermore,  $P$ \textbf{subrequires} interface
$i$ if for some
$j$, $j\sqsubseteq i$ and $P$ requires $j$, and
$P$ \textbf{properly} subrequires
$i$  if for some
$j$, $j\sqsubseteq i$, $j\neq i$ and $P$ requires $j$.
\end{definition}

For example,
\[\jum3; \texttt{b2.set:false};\ter;\ter
\quad\begin{cases}
\text{requires}&\texttt{b2.set:false},\\
\text{subrequires}&\texttt{b2.set:false}+i\text{ for
any interface $i$}.
\end{cases}\]

For threads we can consider thread components as pairs 
$(i,T)$. The definitions of \emph{$T$ requires $i$} and 
\emph{$T$ subrequires $i$} are obvious:

\begin{definition}
Let $T$ be some thread with actions in FMN.
Then
$T$ \textbf{requires} interface
$i$ if each element of $i$
occurs in $T$ and conversely, all actions 
occurring in $T$ are elements of $i$.

Furthermore, $T$ \textbf{subrequires} 
interface $i$ if for some
$j$, $j\sqsubseteq i$ and $T$ requires $j$, and
$T$ \textbf{properly} subrequires
$i$  if for some
$j$, $j\sqsubseteq i$, $j\neq i$ and $T$ requires $j$.
\end{definition}
As an example,
$|\jum3; \texttt{b2.set:false};\ter;\ter|$
(sub)requires $\delta$.
Using these last two definitions, we can refine the
(sub)-requirement notions as follows.

\begin{definition}
Let $i$ be an interface and $P$ an inseq. 
Then
$P$ \textbf{$\mathbf 1$-requires} $i$ if $|P|$ requires
$i$,
$P$ \textbf{sub-$\mathbf 1$-requires} $i$ if $|P|$ subrequires
$i$,
$P$ \textbf{(sub-)n-requires} $i$ if $|\jum n;P|$ (sub)requires
$i$,
and
$P$ \textbf{(sub-)\textup{($\mathbf 1,\textit n$)}-requires} $i$ if for all $m$
with $1\leq m\leq n$,
$P${(sub-)$m$-requires} $i$.
\end{definition}
For example,
\[\jum3; \texttt{b2.set:false};\ter;\ter
\quad\text{(sub)-1-requires }\delta,\quad\text{but }
\text{(sub)-2-requires }\texttt{b2.set:false}.\]

Let $c=(i,P)$ be an instruction sequence component.
By default we assume that $P$ subrequires $i$ unless
explicitly stated otherwise. 
We further call $i$ the interface of $c$ and $P$
the body of $c$.
So for an instruction sequence component $c$ it may
for instance be the case that its body
(sub-)$(1,7)$-requires its interface.
Here are some elementary connections:
\begin{itemize}
\item
if $P$ subrequires $i$ then for all $n$,
$P$ sub-$n$-requires $i$,
\item conversely, if for all $n$,
$P$ \emph{sub-$n$-requires} $i$ then
$P$ subrequires $i$,
\item if for $m=1,...,\length(P)$, where $\length(P)$
is the length of $P$ (its number of instructions), $P$
$m$-requires $i_m$ then $P$ requires
$+_{m=1}^{\length(P)} i_m$.
\end{itemize}

Further decoration of an inseq component with for instance 
information about its inseq notation is possible but will
not be considered here.

\section{Service Components}
\label{sec:serv}
For a service $H$ we assume that it offers processing
for a number of methods collected in a method interface.
Thread-service composition is briefly explained in 
\cite{PZ06}; inseq-service composition 
was introduced in \cite{BP02} (where services were 
called `state machines') and defined as the
composition obtained after thread extraction of the
program term involved. The definition of a service
$H$ includes that of its \emph{reply
function}, the function that determines
the reply to a call of any of its methods (which
in general is based on initialization assumptions
and on its history, i.e., the sequence
of earlier calls). Furthermore, there is a so-called
\emph{empty}
service $\emptyset$ that has the empty method interface
$\delta$.

A method interface (MI)
is either $\delta$ or is created by + and 
$\cdot$ from constants in $V_{LDC}$.

\begin{example}
\label{stack}\rm
A stack over values \texttt{0}, \texttt{1} and \texttt{2}
can be defined as a service with methods \texttt{push:$i$},
\texttt{topeq:$i$} ,
and \texttt{pop}, for $i = \texttt 0,..., \texttt 2$,
where \texttt{push:$i$} pushes $i$ onto the stack and
yields \tr, the action \texttt{topeq:$i$} tests whether 
$i$ is on top of the stack, and \texttt{pop}
pops the stack with reply \tr\ if it is non-empty, 
and otherwise does nothing and yields \fa.

With $\beta$ ranging over $(V_{LDC})^*$,
the reply function $F$ of the stack is 
informally defined by
\begin{align*}
F(\beta\:\texttt{push:$i$})&=\tr 
\quad\text{and generates a stack with $i$ on top}\\
F(\beta\:\texttt{pop})&=\begin{cases}
\tr&\text{if the stack generated by $\beta$ is non-empty}\\
&\text{and then pops this stack},\\
\fa&\text{otherwise,}
\end{cases}\\
F(\beta\:\texttt{topeq:$i$})&=\begin{cases}
\tr&\text{if the stack generated by $\beta$ has $i$ on top}\\
\fa&\text{otherwise.}
\end{cases}
\end{align*}
For a formal definition of a stack as a service, 
see e.g.\ \cite{PZ06}.
\end{example}

\begin{definition}
A service $H$ \textbf{provides} interface
$i$ if 
$H$ offers a 
reply function for all elements 
of $i$, and $H$ \textbf{superprovides} $i$ if for some
$j\sqsupseteq i$, $H$ provides $j$.
\end{definition}

For a \emph{service component} $(j,H)$ we assume that $H$ 
superprovides the method interface $j$ unless
explicitly stated otherwise. 

Below we consider two forms of inseq-service
composition: the \emph{use} and the \emph{apply} composition.
We first discuss the use operator. For an inseq $P$
a use application is written
\[P/_\beta\: H\]
where $\beta$ is a focus. A use application always yields 
a thread, which is reasonable because it is upon
the execution of an inseq that its generated actions may 
call for some service and use its reply for subsequent
execution. The use operator simply drops the service
upon termination or deadlock.  Use
compositions are defined in \cite{BP02} and take a thread
as their left argument, but inseqs can be used instead by 
setting $P/_\beta\: H=|P|/_\beta\: H$.
Below we lift use compositions to the level of components.

\begin{example}
\label{stack2}
\rm
The stack described in Example~\ref{stack} provides
the method interface $j$ defined by
\begin{equation}
\label{j}
j=(\texttt{push:}+\texttt{topeq:})(\texttt 0+\texttt 1+\texttt 2)+\texttt{pop}.
\end{equation}

Let $P$ be a program defined in PGLB$_{i+\texttt{c.}j}$ 
for the FMI $i$ defined in Example~\ref{int}.
Then $P$ can use the stack defined in 
Example~\ref{stack} via focus \texttt c. 
If we only push the value \texttt 0
(so the stack behaves as a counter), we can write $C(n)$
for a stack
holding $n$ times the value \texttt 0 (so $C(0)$
represents the empty stack). 
With the defining equations from \cite{BP02}
it follows that
\begin{align*}
(\texttt{c.push:0};P) 
/_{\texttt c}\: C(n) &= P/_{\texttt c}\: C(n+1),\\
(-\texttt{c.pop};\ter;P)
/_{\texttt c}\: C(0) &= \st,\\
(-\texttt{c.pop};\ter;P) 
/_{\texttt c}\: C(n+1) &= P/_{\texttt c}\:{\texttt c}\:C(n).
\end{align*}
Instructions in PGLB$_{i+\texttt{c.}j}$ 
with foci different from \texttt c
are distributed over use applications, e.g.,
\[(+\texttt{g.set:true};\texttt{f.get};P)/_{\texttt c}\:C(n)=
((\texttt{f.get};P)/_{\texttt c}\:C(n))\unlhd
\texttt{g.set:true}\unrhd
(P/_{\texttt c}\:C(n)).\]
\end{example}

\begin{definition}
For an instruction sequence component
$(i,P)$,
a service component $(j,H)$, and 
some $\betaw\in V_{LDC}^+$,
the \textbf{use composition}
\[(i,P)/_\betaw\: (j,H)\]
is \textbf{matching} if
$\dd{\betaw\texttt .}(i)\sqsubseteq j\quad\text{(note the period that occurs
in $\dd{\betaw\texttt .}$)},$
in which case
\[(i,P)/_\betaw\: (j,H)=\big(\ddm{\betaw\texttt .}(i),~P/_\betaw\: H\big).\]
The  {use composition}
$(i,P)/_\betaw\: (j,H)$ is \textbf{non-matching} if
$\dd{\betaw\texttt .}(i)\not\sqsubseteq j$
and then
\[(i,P)/_\betaw\: (j,H)=\big(\delta,~\di).\]
\end{definition}

\begin{example}[Example~\ref{stack2} continued]
\label{stack3}
\rm
We find for inseq component 
$(i+\texttt{c.}j,~\texttt{c.push:0};P) $
that
\begin{align*}
(i+\texttt{c.}j,~\texttt{c.push:0};P) 
/_{\texttt c}\: (j,C(n))&= (i,~P/_{\texttt c}\: C(n+1),
\end{align*}
while $(i,~\texttt{c.push:0};P) 
/_{\texttt c}\: (j,C(n))=(\delta,\di)$.
\end{example}
\bigskip

The \emph{apply composition} of a thread $T$
and a service $H$ is written
\[T\bullet_\beta H,\]
and always yields a service. Thus threads are 
used to alter the state of a particular service,
and only finite threads without deadlock ($\di$) do 
this in a meaningful way. 
Apply compositions are defined in \cite{BP02} and 
take a thread as left-argument, and a typical
defining equation is 
\[\st\bullet_\beta H=H.\]
However, we can use inseqs 
as left-arguments by 
defining $P\bullet_\beta\: H=|P|\bullet_\beta\: H$.
Below we lift apply compositions to the level of components.

\begin{definition}
For an instruction sequence component
$(i,P)$,
a service component $(j,H)$, and 
some $\betaw\in V_{LDC}^+$,
the \textbf{apply composition}
\[(i,P)\bullet_\betaw (j,H)\]
is \textbf{matching} if
$i\sqsubseteq \betaw\texttt . j$
and then
\[(i,P)\bullet_\betaw (j,H)=\big(j,~P\bullet_\betaw H\big).\]
The  {apply composition}
$(i,P)\bullet_\betaw\: (j,H)$ is \textbf{non-matching} if
$i\not\sqsubseteq \betaw\texttt .j$
and then
\[(i,P)\bullet_\betaw\: (j,H)=\big(\delta,~\emptyset)\]
where 
$\emptyset$ is the empty service.
\end{definition}

\begin{example}\rm
For the stack and the interfaces $i$ and $j$ defined in 
Examples~\ref{stack}--\ref{stack3},
we find
\[(\texttt{c.}j,~\texttt{c.push:0};\texttt{c.push:0};\ter) 
\bullet_{\texttt c}\: (j,C(n))=(j,C(n+2)),\]
while $(i,~\texttt{c.push:0};
\texttt{c.push:0};\ter) 
\bullet_{\texttt c}\: (j,C(n))=(\delta,~\emptyset)$.
\end{example}

\section{Plots and Related Work}
\label{sec:conc}
By allowing interfaces which are not mimimal (subrequired) more options
arise for finding concise interface notations. One might question the
plausibility of components with interfaces which are larger than strictly
necessary. Here are some plots that might lead to that situation.

\begin{enumerate}
\item $(i,P)$ with $P$ the result of complex projections
that are not yet computed.
\item 
$(i,P)$ with $P$ projected to PGLB in fragments (JIT 
projection). At no point in time the complete interface
of $P$ is known.
\item
$(i,P)$ claims name space while preventing redesign
of relevant methods of the services used. This creates 
degrees of freedom for redesign of $P$. 
For example, if $P$ uses booleans 
$\texttt{b1},...,\texttt{b50}$, one
can reserve $\texttt{b1},...,\texttt{b100}$ 
to facilitate future ideas to
optimize $P$.
\item
$(i,P)$ is one of a series inseq components that one 
may want to plug in 
in an execution architecture \cite{BP07}.
The interface $i$ is kept simple to facilitate
a quick check. 
\end{enumerate}
Without any doubt, more plots that justify the existence 
of inseq components can be thought of. However, as stated
earlier, the main
argument for the introduction of (right) progression rings 
introduced in
this paper is to provide a flexible and practical notation for
interface specification by means of an appropriate abstract 
datatype specification.

In \cite{BM07}, a process component is taken as
a pair of an interface and a process specifiable in the 
process algebra ACP \cite{BW90,F00}. In that
paper interfaces are formalized by means of an 
\emph{interface group}, which allows for the distinction 
between expectations and promises in interfaces of process 
components, a distinction that
comes into play in case components with both client and server
behaviour are involved.
However, in our approach the interaction between instruction 
sequences (or threads) and services is not sufficiently symmetric
to use the group structure.

\bibliographystyle{alpha}

\end{document}